\documentclass[sigconf]{acmart}

\usepackage{booktabs} 
\usepackage{balance}

\begin{document}
\title{Further Study on GFR Features for JPEG Steganalysis}

\author{Chao Xia}
\affiliation{%
  \institution{$^{1}$State Key Laboratory of Information Security, Institute of Information Engineering, Chinese Academy of Sciences,}
  \city{Beijing}
  \state{China}
  \postcode{100093}
}
\affiliation{%
  \institution{$^{2}$School of Cyber Security, University of Chinese Academy of Sciences,}
  \city{Beijing}
  \state{China}
  \postcode{100093}
}
\email{xiachao@iie.ac.cn}
\author{Qingxiao Guan}
\authornote{Corresponding author}
\affiliation{%
  \institution{$^{1}$State Key Laboratory of Information Security, Institute of Information Engineering, Chinese Academy of Sciences,}
  \city{Beijing}
  \state{China}
  \postcode{100093}
}
\affiliation{%
  \institution{$^{2}$School of Cyber Security, University of Chinese Academy of Sciences,}
  \city{Beijing}
  \state{China}
  \postcode{100093}
}
\email{guanqingxiao@iie.ac.cn}
\author{Xianfeng Zhao}
\affiliation{%
  \institution{$^{1}$State Key Laboratory of Information Security, Institute of Information Engineering, Chinese Academy of Sciences,}
  \city{Beijing}
  \state{China}
  \postcode{100093}
}
\affiliation{%
  \institution{$^{2}$School of Cyber Security, University of Chinese Academy of Sciences,}
  \city{Beijing}
  \state{China}
  \postcode{100093}
}
\email{zhaoxianfeng@iie.ac.cn}

\begin{abstract}
The GFR (Gabor Filter Residual) features,
built as histograms of quantized residuals obtained with 2D Gabor filters, can achieve competitive detection performance against adaptive JPEG steganography.
In this paper, an improved version of the GFR is proposed.
First,
a novel histogram merging method is proposed according to the symmetries between different Gabor filters, thus making the features more compact and robust.
Second, a new weighted histogram method is proposed by considering the position of the residual value in a quantization interval, making the features more sensitive to the slight changes in residual values.
The experiments are given to demonstrate the effectiveness of our proposed methods.
Finally, we design a CNN to duplicate the detector with the improved GFR features and the ensemble classifier,
thus optimizing the design of the filters used to form residuals in JPEG-phase-aware features.
\end{abstract}

%
%





\maketitle

\section{Introduction}
The purpose of steganography is to embed secret messages into cover objects without arousing a warder's suspicion.
Steganalysis, the counterpart of steganography, aims to detect the presence of hidden data.
Since JPEG is widely used in modern society, especially in the Internet communication, much attention has been attached to this ideal cover.
With the advent of the STCs (Syndrome-Trellis Codes) coding technique \cite{STC},
some adaptive JPEG steganographic methods have been designed in recent years,
such as UED (Uniform Embedding Distortion)~\cite{UED} and J-UNIWARD (JPEG Universal Wavelet Relative Distortion)~\cite{UNIWARD}.
These adaptive methods are difficult to detect because the embedding changes are localized in complex
content which is hard to model.

To attack adaptive JPEG steganography well,
the DCTR (Discrete Cosine Transform Residual)~\cite{DCTR} opens up a new framework of JPEG phase-aware features.
The DCTR, using the histograms of residuals obtained with 64 DCT kernels, not only has relatively low complexity but also provides good detection performance.
In~\cite{PHARM}, the PHARM (Phase-Aware Projection Model), following this phase-aware framework,
computes the histograms of multiple random projections of residuals obtained with linear pixel predictors.
Random projections diversify the model in a similar manner as in the PSRM (Projection Spatial Rich Model)~\cite{PSRM}, improving the detection accuracy further.
There are three important observations in the design of the DCTR and the PHARM.
\textbf{First}, unlike the previous JPEG steganalysis feature sets (e.g., PEV~\cite{PEV}, JRM~\cite{JRM}),
both the DCTR and the PHARM are constructed in the spatial domain rather than the JPEG domain.
Before obtaining noise residuals, JPEG images are decompressed to the spatial domain without rounding to integers.
This is probably because the statistical characteristics captured in the spatial domain are more sensitive to
adaptive JPEG embedding algorithms~\cite{WHYS}.
\textbf{Second}, phase-awareness is employed in these two feature sets.
Instead of directly computing the histogram features from all values of the whole residual,
both feature sets compute the histograms from 64 subsets of the residual,
for the statistical properties of pixels in a decompressed JPEG image differ w.r.t. their positions within the $8\times8$ pixel grid.
\textbf{Third}, symmetrization is useful for forming the final features.
The symmetries are utilized to reduce the feature dimension and make them more robust.
The GFR (Gabor Filter Residual)~\cite{GFR} is motivated by these three observations. The difference is that the GFR uses the histograms of residuals obtained using 2D Gabor filters.
The 2D Gabor filters can describe image texture features from different scales and orientations.
Thus, the GFR can achieve the state-of-the-art performance in most of the cases when steganalyzing adaptive JPEG steganography.

In this paper, we revisit the design of the GFR and attempt to further improve its performance. The main contributions can be concluded as follows.
\textbf{First}, a new histogram merging method is proposed. In the GFR, the histograms computed from 64 subsets of one Gabor residual are merged with the method designed for the DCTR. But this strategy is not proper,
for the symmetric properties of the Gabor filters differ from the DCT filters. Thus, we merge the histograms of one Gabor residual in a different way. Then,  according to the symmetries between Gabor
filters, histograms of different Gabor residuals are merged further to make the final features more compact and powerful.
\textbf{Second}, a novel weighted histogram method is proposed. In the GFR, histograms are computed from quantized residuals.
Although the quantization is meaningful for steganalysis,
it may inevitably lose part of useful information.
With the quantization, the histograms in the GFR can only reflect the changes that enable the residual values to shift from a quantization interval to another,
while leaving out those small changes.
To avoid this situation, we propose a novel way to compute the histograms using a weighted voting scheme without a rounding operation.
This scheme takes into account the small disturbance of residual values within a quantization interval, thus making the histogram features more effective.
\textbf{Third}, a novel CNN architecture, with proper initialization, is elaborated to duplicate the steganalytic scheme with the improved GFR features and FLD-ensemble. Within our network, the kernels in the convolutional layer
are updated during the training, showing the potential to obtain the filters which are more suitable for forming residuals in JPEG-phase-aware steganalysis features.

In this paper, we call the new feature set the GFR-GW (GFR-Gabor symmetric merging and Weighted histograms) which applies the proposed histogram merging method and our weighted histogram method. And the histogram features only using the proposed merging method are called the GFR-GSM (GFR-Gabor Symmetric Merging) features. The experimental results will be given to show the advantages of the proposed features in the detection of adaptive JPEG steganography.
The rest of this paper is organized as follows. In Section~\ref{Sec_2}, we describe the original GFR features briefly.
In Section~\ref{Sec_3}, we discuss the reason why the histograms of 64 subsets of one Gabor residual can not be merged with the same method in the DCTR.
In Section~\ref{Sec_4},
based on the symmetries between Gabor filters, we propose our method to merge the histograms of the subsets of different Gabor residuals.
In Section~\ref{Sec_5}, our weighted voting scheme for histogram computation is introduced.
In Section~\ref{Sec_6}, the proposed features (the GFR-GSM and the GFR-GW) are compared with other JPEG steganalysis features by experiments.
In Section~\ref{Sec_7}, a novel CNN is proposed to duplicate the scheme with GFR-GW and FLD ensemble classifier.
Conclusions and future work are given in Section~\ref{Sec_8}.

\section{Original GFR Features}
\label{Sec_2}
The GFR features compute the histograms from the subsets of residuals obtained using 2D Gabor filters.
The 2D Gabor filters help the GFR to capture the effect of the steganography in different scales and orientations.
In this section, we briefly describe how to calculate the original GFR features to make this paper self-contained.
We do not go into the details which can be seen in the original literature~\cite{GFR}.

For the GFR, the calculation procedures are described as follows.

\textbf{Step 1}: A JPEG image is decompressed to the spatial domain without rounding the pixel values to the discrete set \{0,1, \ldots , 255\},
i.e., the gray values of pixels are preserved in the form of real numbers.

\textbf{Step 2}: The 2D Gabor filter bank is generated and the bank in~\cite{GFR}
includes 2D Gabor filters with 2 phase offsets ($\phi=0, \pi$), 4 scales ($\sigma=0.5, 0.75, 1, 1.25$) and 32 orientations ($\theta=0, \pi/32, \ldots, 31\pi/32$).

\textbf{Step 3}: The decompressed JPEG image is convolved with the $8 \times 8$ 2D Gabor filter $\mathbf{G}^{\phi,\sigma,\theta}$
to get the corresponding residual image $\mathbf{U}^{\phi,\sigma,\theta}$.

\textbf{Step 4}: According to the JPEG phase ($a,b$) ($0 \leq a,b \leq 7$),
the residual $\mathbf{U}^{\phi,\sigma,\theta}$ is divided into 64 subsets $\mathbf{U}^{\phi,\sigma,\theta}_{a,b}$ by interval 8 down-sampling.

\textbf{Step 5}: The histogram feature $\mathbf{h}^{\phi,\sigma,\theta}_{a,b}$ is computed from each subset $\mathbf{U}^{\phi,\sigma,\theta}_{a,b}$.

\begin{equation}
\mathbf{h}^{\phi,\sigma,\theta}_{a,b}(r) = \frac{1}{\left|\mathbf{U}^{\phi,\sigma,\theta}_{a,b}\right|}\sum_{u \in \mathbf{U}^{\phi,\sigma,\theta}_{a,b}} [Q_{T}(|u|/q)=r],
\end{equation}
where $Q_{T}$ is a quantizer quantizing the residual samples to integer centroids \{0, 1, \ldots, T\}, $q$ is the quantization step,
and $[P]$ is the Iverson bracket equal to 1 when statement $P$ is true and 0 when $P$ is false.

\textbf{Step 6}:
For residual $\mathbf{U}^{\phi,\sigma,\theta}$,
all the 64 histograms $\mathbf{h}^{\phi,\sigma,\theta}_{a,b}$
are merged into 25 according to the same method in the DCTR~\cite{DCTR}.
Then these 25 histograms are concatenated to obtain the histogram feature $\mathbf{h}^{\phi,\sigma,\theta}$
of residual $\mathbf{U}^{\phi,\sigma,\theta}$.

\textbf{Step 7}:
The histogram features $\mathbf{h}^{\phi,\sigma,\pi-\theta}$ and $\mathbf{h}^{\phi,\sigma,\theta}$ are merged together
according to the symmetric orientations.

\textbf{Step 8}:
All the merged histograms are concatenated to form the GFR features.

\section{Difference between Gabor filters and DCT filters}
\label{Sec_3}
From the description of the GFR, it can be seen that there are two steps in merging histograms in the GFR.
First, in \textbf{Step 6}, the histograms of 64 subsets of one Gabor residual are merged together.
Second, in \textbf{Step 7}, we merge the histograms of two residuals
with symmetric directions.
In this section, we discuss \textbf{Step 6}, where
the 64 histograms $\mathbf{h}^{\phi,\sigma,\theta}_{a,b}$ are merged in the same manner as in the DCTR
where the residuals are obtained using the DCT filters.
In the DCTR, 64 histograms computed from  64 subsets of one DCT residual are merged into 25
according to the symmetries of the projection vectors of DCTR.
However, the symmetric properties of the Gabor filters differ from the DCT filters,
which leads to different kinds of the symmetries of the projection vectors of GFR.
Hence, it is more reasonable to merge the histograms $\mathbf{h}^{\phi,\sigma,\theta}_{a,b}$ in a different way rather than in \textbf{Step 6} of the GFR.

In this section, we first introduce the symmetric properties of the DCT filters and the Gabor filters respectively
and show the difference between them.
After describing the merging method in the DCTR,
we discuss how to merge the histograms of 64 subsets of one Gabor residual.

In this paper, the DCT filter is denoted as $\mathbf{B}^{i,j}$, where $i$, $j$ indicate the spatial frequencies,
and $0\leq i,j \leq7$.
The Gabor filter is denoted as $\mathbf{G}^{\phi,\sigma,\theta}$,
where $\theta$ is the orientation parameter,
$\sigma$ is the scale parameter and $\phi$ is the phase shift.


\subsection{Symmetric Properties of DCT Filters and Gabor Filters}
\label{Sec_3_1}
The symmetric properties of filters are related to the symmetries of the projection vectors.
Therefore, we first introduce the symmetric properties of the DCT filters and the Gabor filters, respectively.

For the DCT filter $\mathbf{B}^{i,j}$ ($0\leq i,j \leq7$),
it is symmetric or antisymmetric in either direction:
\begin{equation}
\label{eq_symmetry_DCT}
\mathbf{B}^{i,j} = \left\{
\begin{aligned}
&\mathrm{flipud}(\mathbf{B}^{i,j}) & & i \ \ \mathrm{is \ even} \\
-&\mathrm{flipud}(\mathbf{B}^{i,j}) & & i \ \ \mathrm{is \ odd} \\
&\mathrm{fliplr}(\mathbf{B}^{i,j}) & & j \ \ \mathrm{is \ even} \\
-&\mathrm{fliplr}(\mathbf{B}^{i,j}) & & j \ \ \mathrm{is \ odd}
\end{aligned}\right.,
\end{equation}
where $\mathrm{flipud}(\cdot)$ denotes the flipping operator that flips a matrix vertically and
$\mathrm{fliplr}(\cdot)$ denotes the operator that flips a matrix horizontally.

For the Gabor filter $\mathbf{G}^{\phi,\sigma,\theta}$,
both in~\cite{GFR} and in this paper,
the phase shift $\phi$ is set as $0$ and ${\pi}/{2}$. Then, we have
\begin{equation}
\mathbf{G}^{\phi,\sigma,\pi+\theta} = - \mathbf{G}^{\phi,\sigma,\theta}, \ 0 \leq \theta < \pi.
\end{equation}
The absolute values of residual images generated by convolving with $\mathbf{G}^{\phi,\sigma,\theta}$
are the same as those with $\mathbf{G}^{\phi,\sigma,\pi+\theta}$.
Thus, we only consider the condition of $0 \leq \theta < \pi$ and select the same 32 orientations
($\theta=0, \pi/32, \ldots, 31\pi/32$) as in the original GFR~\cite{GFR}.

Now we examine the symmetric properties of the Gabor filters $\mathbf{G}^{\phi,\sigma,\theta}$ ($0 \leq \theta < \pi $).
When $\theta = \{0,\pi/2\}$, the Gabor filter $\mathbf{G}^{0,\sigma,\theta = \{0,\pi/2\}}$ is symmetric in both directions,
and the Gabor filter $\mathbf{G}^{\pi/2,\sigma,\theta = \{0,\pi/2\}}$ is symmetric in one direction
and antisymmetric in the other direction:
\begin{equation}
\label{eq_symmetry_Gabor_1}
\begin{aligned}
&\mathbf{G}^{0,\sigma,0}& &=& &\mathrm{flipud}\left(\mathbf{G}^{0,\sigma,0}\right)&
&=&&\mathrm{fliplr}\left(\mathbf{G}^{0,\sigma,0}\right)\\
&\mathbf{G}^{0,\sigma,\frac{\pi}{2}}& &=& &\mathrm{flipud}\left(\mathbf{G}^{0,\sigma,\frac{\pi}{2}}\right)&
&=&&\mathrm{fliplr}\left(\mathbf{G}^{0,\sigma,\frac{\pi}{2}}\right)\\
&\mathbf{G}^{\frac{\pi}{2},\sigma,0}& &=& &\mathrm{flipud}\left(\mathbf{G}^{\frac{\pi}{2},\sigma,0}\right)&
&=& -&\mathrm{fliplr}\left(\mathbf{G}^{\frac{\pi}{2},\sigma,0}\right)\\
&\mathbf{G}^{\frac{\pi}{2},\sigma,\frac{\pi}{2}}& &=& -&\mathrm{flipud}\left(\mathbf{G}^{\frac{\pi}{2},\sigma,\frac{\pi}{2}}\right)&
&=& &\mathrm{fliplr}\left(\mathbf{G}^{\frac{\pi}{2},\sigma,\frac{\pi}{2}}\right)
\end{aligned}.
\end{equation}
However,  when $\theta \neq \{0,{\pi}/{2}\}$, unlike DCT filters,
$\mathbf{G}^{\phi,\sigma,\theta \neq \{0,{\pi}/{2}\}}$
is neither symmetric nor antisymmetric in any direction.
But $\mathbf{G}^{\phi,\sigma,\theta \neq \{0,{\pi}/{2}\}}$ is centrosymmetric or anti-centrosymmetric.
When $\phi = {0}$, the Gabor filter $\mathbf{G}^{0,\sigma,\theta \neq \{0,{\pi}/{2}\}}$ is centrosymmetric,
and when $\phi = {\pi}/{2}$,
the Gabor filter $\mathbf{G}^{{\pi}/{2},\sigma,\theta \neq \{0,{\pi}/{2}\}}$ is anti-centrosymmetric:
\begin{equation}
\label{eq_symmetry_Gabor_2}
\begin{aligned}
\end{aligned}
\begin{aligned}
&\forall \phi, \  \sigma, \  \theta \neq 0,\frac{\pi}{2}
\\
&\mathbf{G}^{\phi,\sigma,\theta} \neq \pm \ \mathrm{flipud}\left(\mathbf{G}^{\phi,\sigma,\theta}\right),\
\mathbf{G}^{\phi,\sigma,\theta} \neq \pm \ \mathrm{fliplr}\left(\mathbf{G}^{\phi,\sigma,\theta}\right),
\\
\\
&\forall \sigma, \  \theta \neq 0,\frac{\pi}{2}
\\
&\mathbf{G}^{0,\sigma,\theta} = \mathrm{rot180}\left(\mathbf{G}^{0,\sigma,\theta}\right),\
\mathbf{G}^{\frac{\pi}{2},\sigma,\theta} = -\mathrm{rot180}\left(\mathbf{G}^{\frac{\pi}{2},\sigma,\theta}\right),
\end{aligned}
\end{equation}
where $\mathrm{rot180}(\cdot)$ is a rotation operator that rots the matrix by 180 degrees.


\subsection{Merging Method in the DCTR}
\label{Sec_3_2}
\begin{figure*}
\centering
    \includegraphics[width=2\columnwidth]{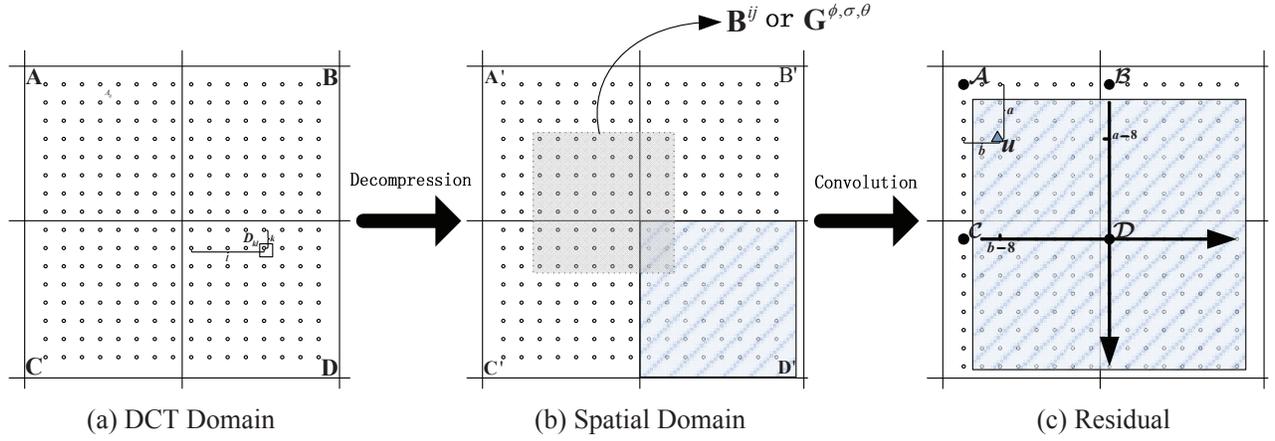}
    \caption{The computing process of a residual image (DCT residual or Gabor residual).
             Left: Dots indicate the DCT coefficients, and $A$, $B$, $C$, $D$ are four neighboring DCT blocks.
             Middle: Dots indicate the pixels in the decompressed JPEG image, and $A'$, $B'$, $C'$, $D'$ are the corresponding pixel blocks.
             Right: Dots indicate the residual samples in the DCT residual or Gabor residual,
             and the element $u$ is generated by convolving 64 pixels in the dotted line block with $\mathbf{B}^{i,j}$ or $\mathbf{G}^{\phi,\sigma,\theta}$.
             The change of the DCT coefficient $D_{kl}$ will affect all $8\times8$ pixels in block $D'$. And a $15\times15$ neighborhood of values in the residual image (in the shaded region) will be modified. The position of the residual sample $\mathcal{D}$ is at the center of the shaded region and the coordinate of the position of $u$ (the triangle) in the shaded region is $(a-8, b-8)$.}
    \label{Figure:DCT2RS}
\end{figure*}

In order to realize the relationship between the symmetric properties of the filters and the method of merging histograms,
we rephrase the merging method in the DCTR, which is also used in the original GFR.
As shown in \textbf{Figure~\ref{Figure:DCT2RS}}, from the computing process of a residual image (DCT residual or Gabor residual), we find that
the modification of one DCT coefficient ($D_{ij}$ in the DCT block $D$ in \textbf{Figure~\ref{Figure:DCT2RS}(a)}) will affect the values of all $8 \times 8$ pixels in the corresponding block in the spatial domain (pixels in the $8 \times 8$ pixel block $D'$ in \textbf{Figure~\ref{Figure:DCT2RS}(b)}) because of the JPEG decompression.
Then the values of $15 \times 15$ residual samples (the shaded region in \textbf{Figure~\ref{Figure:DCT2RS}(c)}) will be changed by convolving with an $8 \times 8$ filter (DCT filter or Gabor filter). Specifically, due to changing one
DCT coefficient, a $15 \times 15$ neighborhood of values in the DCT residual will be modified by
\begin{equation}
\mathbf{R}^{(i,j)(k,l)} = \mathbf{B}^{i,j} \otimes  \mathbf{B}^{k,l},
\end{equation}
where the modified DCT coefficient is in mode $(k,l)$, $\mathbf{B}^{i,j}$ denotes the DCT filter used to convolve the decompressed JPEG image,
and $\otimes$ denotes the full cross-correlation.

According to the symmetric properties of the DCT filters (\ref{eq_symmetry_DCT}),
we can see that when indexing $\mathbf{R}^{(i,j)(k,l)} \in \mathbb{R}^{15 \times 15}$ with indices in
\{ -7, -6, \ldots, -1, 0, 1, \ldots, 6, 7\}, $\mathbf{R}^{(i,j)(k,l)}$ satisfies the following symmetry
\begin{equation}
\label{eq_symmetry_R1}
{R}^{(i,j)(k,l)}_{a,b} = \left\{
\begin{aligned}
{{R}}^{(i,j)(k,l)}_{-a,b} & & &(i+k) \ \ \text{is \ even} \\
-{R}^{(i,j)(k,l)}_{-a,b} & & &(i+k) \ \ \text{is \ odd} \\
{R}^{(i,j)(k,l)}_{a,-b} & & &(j+l) \ \ \text{is \ even} \\
-{R}^{(i,j)(k,l)}_{a,-b} & & &(j+l) \ \ \text{is \ odd}
\end{aligned}\right..
\end{equation}
From the symmetry of $\mathbf{R}^{(i,j)(k,l)}$ (\ref{eq_symmetry_R1}),
we can see that $\left|\mathbf{R}^{(i,j)(k,l)}\right|$ is symmetric about both axes
\begin{equation}
\label{eq_symmetry_|R1|}
\begin{aligned}
\left|{R}^{(i,j)(k,l)}_{a,b}\right|  = & \left|{R}^{(i,j)(k,l)}_{-a,b}\right|\\
\left|{R}^{(i,j)(k,l)}_{a,b}\right|  = & \left|{R}^{(i,j)(k,l)}_{a,-b}\right|
\end{aligned}.
\end{equation}

We now show how to compute a particular value $u$ in the DCT residual
(the location of $u$ is marked by a triangle in \textbf{Figure~\ref{Figure:DCT2RS}(c)}).
In \textbf{Figure~\ref{Figure:DCT2RS}(c)},
four residual samples
$\mathcal{A}$, $\mathcal{B}$, $\mathcal{C}$, $\mathcal{D}$ (black circles in \textbf{Figure~\ref{Figure:DCT2RS}(c)}) are computed by positioning the DCT filter $\mathbf{B}^{i,j}$
within one pixel block (e.g., $\mathcal{D}$ is generated by only convolving $8\times8$ pixels in $D'$ with $\mathbf{B}^{i,j}$).
After decompression and convolution, the effect of the DCT coefficient $D_{kl}$ on the DCT residual can be
expressed as $Q_{kl}D_{kl}\mathbf{R}^{(i,j)(k,l)}$ . The location of $\mathcal{D}$ is at the center of $Q_{kl}D_{kl}\mathbf{R}^{(i,j)(k,l)}$ and the relative position of $u$ w.r.t $\mathcal{D}$ is $(a-8,b-8)$.
Similarly, the relative locations of $u$ w.r.t. the other three centers $\mathcal{A}$, $\mathcal{B}$, $\mathcal{C}$ are $(a,b)$, $(a,b-8)$ and $(a-8,b)$, respectively.
The value $u$ can be calculated as follows:
\begin{equation}
\begin{aligned}
u = &\sum^{7}_{k=0}\sum^{7}_{l=0}Q_{kl}\left[A_{kl}R^{(i,j)(k,l)}_{a,b} + B_{kl}R^{(i,j)(k,l)}_{a,b-8}\right.\\
    &+ \left.C_{kl}R^{(i,j)(k,l)}_{a-8,b} + D_{kl}R^{(i,j)(k,l)}_{a-8,b-8}\right],
\end{aligned}
\end{equation}
where
$A_{kl}$, $B_{kl}$, $C_{kl}$, $D_{kl}$ are the DCT coefficients of the corresponding four neighboring DCT blocks ($A$, $B$, $C$, $D$),
and $Q_{kl}$ is the quantization step of the $(k,l)$th DCT mode.

The value $u$ can also be denoted as a projection of 256 dequantized DCT coefficients from the
four adjacent DCT blocks with a projection vector of DCTR $\mathbf{P}^{i,j}_{a,b}$
\begin{equation}
\label{eq_definition_projection_DCTR}
 u=    \begin{pmatrix}
           Q_{00}A_{00}\\
           \vdots\\
           Q_{00}B_{00}\\
           \vdots\\
           Q_{00}C_{00}\\
           \vdots\\
           Q_{00}D_{00}\\
           \vdots\\
           Q_{77}D_{77}
           \end{pmatrix}^{T}
           \cdot
     \underbrace{
           \begin{pmatrix}
           R^{(i,j)(0,0)}_{a,b}\\
           \vdots\\
           R^{(i,j)(0,0)}_{a,b-8}\\
           \vdots\\
           R^{(i,j)(0,0)}_{a-8,b}\\
           \vdots\\
           R^{(i,j)(0,0)}_{a-8,b-8}\\
           \vdots\\
           R^{(i,j)(7,7)}_{a-8,b-8}
           \end{pmatrix}
           }_{\mathbf{P}^{i,j}_{a,b}}.
\end{equation}
From the symmetry of $\left|\mathbf{R}^{(i,j)(k,l)}\right|$ (\ref{eq_symmetry_|R1|})
and the definition of the projection vector (\ref{eq_definition_projection_DCTR}),
we can see that the absolute values of the projection vector $\left|\mathbf{P}^{i,j}\right|$ follow the symmetry
\begin{equation}
\label{eq_symmetry_|p|_1}
\left|\mathbf{P}^{i,j}_{a,b}\right| = \left|\mathbf{P}^{i,j}_{a,-b}\right|
=\left|\mathbf{P}^{i,j}_{-a,b}\right|
= \left|\mathbf{P}^{i,j}_{-a,-b}\right|.
\end{equation}
Because the size of the DCT block is $8 \times 8$, the projection vectors of DCTR satisfy the following symmetry as described in~\cite{DCTR}
\begin{equation}
\label{eq_symmetry_|p|_2}
\left|\mathbf{P}^{i,j}_{a,b}\right| = \left|\mathbf{P}^{i,j}_{a,b-8}\right| = \left|\mathbf{P}^{i,j}_{a-8,b}\right| = \left|\mathbf{P}^{i,j}_{a-8,b-8}\right|.
\end{equation}

Combining (\ref{eq_symmetry_|p|_1}) and (\ref{eq_symmetry_|p|_2}), we have the symmetry that
is used in the merging method in the DCTR
\begin{equation}
\label{eq_symmetry_|p|_3}
\left|\mathbf{P}^{i,j}_{a,b}\right| = \left|\mathbf{P}^{i,j}_{a,8-b}\right|
= \left|\mathbf{P}^{i,j}_{8-a,b}\right|
= \left|\mathbf{P}^{i,j}_{8-a,8-b}\right|.
\end{equation}
According to (\ref{eq_symmetry_|p|_3}), hence, we can merge the histograms of the subsets corresponding to the positions $(a,b)$, $(8-a,b)$, $(a,8-b)$, $(8-a,8-b)$ in a DCT residual.

\subsection{Merging Histograms of one Gabor Residual}
\label{Sec_3_3}
However, the symmetric properties of the Gabor filters are different from the DCT filters,
which causes the projection vectors of GFR to satisfy another kind of symmetry.
Thus, the histograms $\mathbf{h}^{\phi,\sigma,\theta}_{a,b}$ of 64 subsets of one Gabor residual can be merged in a different way.

When one DCT coefficient is modified,
a $15 \times 15$ neighborhood of values in the Gabor residual will be modified by
\begin{equation}
\mathbf{R}^{(\phi,\sigma,\theta)(k,l)} = \mathbf{G}^{\phi,\sigma,\theta} \otimes  \mathbf{B}^{k,l},
\end{equation}
where the modified DCT coefficient is in mode $(k,l)$, $\mathbf{G}^{\phi,\sigma,\theta}$ denotes the Gabor filter used to convolve the decompressed JPEG image,
and $\otimes$ denotes the full cross-correlation.

According to the symmetric properties of the Gabor filters (\ref{eq_symmetry_Gabor_1}) and (\ref{eq_symmetry_Gabor_2}) described in Section~\ref{Sec_3_1},
we find that the symmetric properties of $\left|\mathbf{R}^{(\phi,\sigma,\theta)(k,l)}\right|$
depend on the value of the parameter $\theta$.
When $\theta=\{ 0,\pi/2 \}$, $\left|\mathbf{R}^{(\phi,\sigma,\theta)(k,l)}\right|$ satisfies the same symmetry as $\left|\mathbf{R}^{(i,j)(k,l)}\right|$ in the DCTR. That is,
\begin{equation}
\label{eq_symmetry_|R2_1|}
\left|{R}^{(\phi,\sigma,\theta)(k,l)}_{a,b}\right|
=\left|{R}^{(\phi,\sigma,\theta)(k,l)}_{-a,b}\right|
=\left|{R}^{(\phi,\sigma,\theta)(k,l)}_{a,-b}\right|
=\left|{R}^{(\phi,\sigma,\theta)(k,l)}_{-a,-b}\right|.
\end{equation}
However, when $\theta \neq \{ 0,\pi/2 \}$, $\left|\mathbf{R}^{(\phi,\sigma,\theta)(k,l)}\right|$ only satisfies the centrosymmetry
\begin{equation}
\label{eq_symmetry_|R2_2|}
\begin{aligned}
\left|{R}^{(\phi,\sigma,\theta)(k,l)}_{a,b}\right|
& =\left|{R}^{(\phi,\sigma,\theta)(k,l)}_{-a,-b}\right|\\
& \neq\left|{R}^{(\phi,\sigma,\theta)(k,l)}_{-a,b}\right|\\
& \neq\left|{R}^{(\phi,\sigma,\theta)(k,l)}_{a,-b}\right|
\end{aligned}.
\end{equation}

For the GFR, a particular value $u$ in the Gabor residual $\mathbf{U}^{\phi,\sigma,\theta}$ can be computed as follows:
\begin{equation}
\begin{aligned}
u = &\sum^{7}_{k=0}\sum^{7}_{l=0}Q_{kl}\left[A_{kl}R^{(\phi,\sigma,\theta)(k,l)}_{a,b} + B_{kl}R^{(\phi,\sigma,\theta)(k,l)}_{a,b-8}\right.\\
    &+ \left.C_{kl}R^{(\phi,\sigma,\theta)(k,l)}_{a-8,b} + D_{kl}R^{(\phi,\sigma,\theta)(k,l)}_{a-8,b-8}\right].
\end{aligned}
\end{equation}
That is,
\begin{equation}
\label{eq_definition_projection_GFR}
 u=    \begin{pmatrix}
           Q_{00}A_{00}\\
           \vdots\\
           Q_{00}B_{00}\\
           \vdots\\
           Q_{00}C_{00}\\
           \vdots\\
           Q_{00}D_{00}\\
           \vdots\\
           Q_{77}D_{77}
           \end{pmatrix}^{T}
           \cdot
     \underbrace{
           \begin{pmatrix}
           R^{(\phi,\sigma,\theta)(0,0)}_{a,b}\\
           \vdots\\
           R^{(\phi,\sigma,\theta)(0,0)}_{a,b-8}\\
           \vdots\\
           R^{(\phi,\sigma,\theta)(0,0)}_{a-8,b}\\
           \vdots\\
           R^{(\phi,\sigma,\theta)(0,0)}_{a-8,b-8}\\
           \vdots\\
           R^{(\phi,\sigma,\theta)(7,7)}_{a-8,b-8}
           \end{pmatrix}
           }_{\mathbf{P}^{\phi,\sigma,\theta}_{a,b}},
\end{equation}
where $\mathbf{P}^{\phi,\sigma,\theta}_{a,b}$ is a projection vector of GFR.

From the symmetry of $\left|\mathbf{R}^{(\phi,\sigma,\theta)(k,l)}\right|$
(\ref{eq_symmetry_|R2_1|}), (\ref{eq_symmetry_|R2_2|})
and the definition of the projection vector of GFR (\ref{eq_definition_projection_GFR}),
it can be seen that $\left|\mathbf{P}^{\phi,\sigma,\theta}\right|$ follows the symmetry:

$ \ \forall \phi, \ \sigma, \ \theta \in \{0,\pi/2\}$
\begin{equation}
\label{eq_symmetry_|P2_1|}
\left|\mathbf{P}^{\phi,\sigma,\theta}_{a,b}\right|
=\left|\mathbf{P}^{\phi,\sigma,\theta}_{-a,b}\right|
=\left|\mathbf{P}^{\phi,\sigma,\theta}_{a,-b}\right|
=\left|\mathbf{P}^{\phi,\sigma,\theta}_{-a,-b}\right|;
\end{equation}

$ \ \forall \phi, \ \sigma, \ \theta \neq 0,\pi/2$
\begin{equation}
\label{eq_symmetry_|P2_2|}
\begin{aligned}
\left|\mathbf{P}^{\phi,\sigma,\theta}_{a,b}\right|
& =\left|\mathbf{P}^{\phi,\sigma,\theta}_{-a,-b}\right|\\
& \neq\left|\mathbf{P}^{\phi,\sigma,\theta}_{-a,b}\right|\\
& \neq\left|\mathbf{P}^{\phi,\sigma,\theta}_{a,-b}\right|
\end{aligned}.
\end{equation}
The projection vectors of GFR also satisfy the following symmetry
\begin{equation}
\label{eq_symmetry_|P2|_2}
\left|\mathbf{P}^{\phi,\sigma,\theta}_{a,b}\right| = \left|\mathbf{P}^{\phi,\sigma,\theta}_{a,b-8}\right| = \left|\mathbf{P}^{\phi,\sigma,\theta}_{a-8,b}\right| = \left|\mathbf{P}^{\phi,\sigma,\theta}_{a-8,b-8}\right|.
\end{equation}

From (\ref{eq_symmetry_|P2_1|}) and (\ref{eq_symmetry_|P2|_2}), we find that when $\theta=\{ 0,\pi/2 \}$,
the projection vectors of GFR $\left|\mathbf{P}^{\phi,\sigma,\theta}\right|$ satisfy the same symmetry as  $\left|\mathbf{P}^{i,j}\right|$ in the DCTR,
\begin{equation}
\label{eq_symmetry_|P2_1|_3}
\left|\mathbf{P}^{\phi,\sigma,\theta}_{a,b}\right| = \left|\mathbf{P}^{\phi,\sigma,\theta}_{a,8-b}\right| = \left|\mathbf{P}^{\phi,\sigma,\theta}_{8-a,b}\right| = \left|\mathbf{P}^{\phi,\sigma,\theta}_{8-a,8-b}\right|.
\end{equation}
Hence, for the residual $\mathbf{U}^{\phi,\sigma,\theta=\{0,\pi/2\}}$ generated with the Gabor filter whose orientation parameter $\theta = 0,\pi/2$,
the histograms of 64 subsets of $\mathbf{U}^{\phi,\sigma,\theta=\{0,\pi/2\}}$ can be merged in the same way as in the DCTR.
We can merge together the histograms of the subsets corresponding to the positions $(a,b)$, $(8-a,b)$, $(a,8-b)$, $(8-a,8-b)$ in $\mathbf{U}^{\phi,\sigma,\theta=\{0,\pi/2\}}$, and 64 histograms can be merged into 25.

However, from (\ref{eq_symmetry_|P2_2|}) and (\ref{eq_symmetry_|P2|_2}), we find that when $\theta \neq \{ 0,\pi/2 \}$,
the projection vectors of GFR $\left|\mathbf{P}^{\phi,\sigma,\theta}\right|$ satisfy a different kind of symmetry than $\left|\mathbf{P}^{i,j}\right|$ in the DCTR,
\begin{equation}
\begin{aligned}
\label{eq_symmetry_|P2_2|_3}
|\mathbf{P}^{\phi,\sigma,\theta}_{a,b}|  = & |\mathbf{P}^{\phi,\sigma,\theta}_{8-a,8-b}|\\
|\mathbf{P}^{\phi,\sigma,\theta}_{a,b}|  \neq & |\mathbf{P}^{\phi,\sigma,\theta}_{8-a,b}|\\
|\mathbf{P}^{\phi,\sigma,\theta}_{a,b}|  \neq & |\mathbf{P}^{\phi,\sigma,\theta}_{a,8-b}|
\end{aligned}.
\end{equation}
Thus,
the histograms of 64 subsets of $\mathbf{U}^{\phi,\sigma,\theta\neq\{0,\pi/2\}}$ can not be merged in the same way as in the DCTR. However,
we can merge the histograms of the subsets corresponding to the positions $(a,b)$, $(8-a,8-b)$ in $\mathbf{U}^{\phi,\sigma,\theta \neq \{0,\pi/2\}}$,
and 64 histograms can be merged into 34.

\section{Proposed Histogram Merging Method}
\label{Sec_4}

\begin{figure*}
\centering
\includegraphics[width=2\columnwidth]{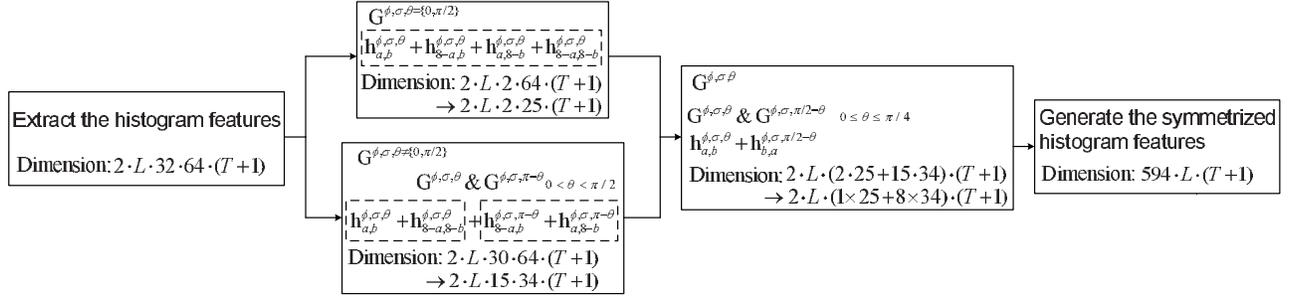}
\caption{The flow of the proposed merging method. The parameter $L$ denotes the number of scales of the Gabor filters, the parameter $T$ means the threshold on residual values, the number of phases of the Gabor filters is 2, the number of orientations of the Gabor filters is 32 and the number of JPEG phases is 64.
}
\label{Figure:flow}
\end{figure*}

In order to further reduce the dimension, we introduce our histogram merging method in this section,
taking into consideration the symmetries between Garbor filters.
As shown in \textbf{Figure~\ref{Figure:flow}}, after merging the 64 histograms $\mathbf{h}^{\phi,\sigma,\theta}_{a,b}$ of one Gabor residual (in the dashed boxes in \textbf{Figure~\ref{Figure:flow}}),
we further merge the histograms of different Gabor residuals
in two steps.

\begin{figure}
\centering
\includegraphics[width=0.9\columnwidth]{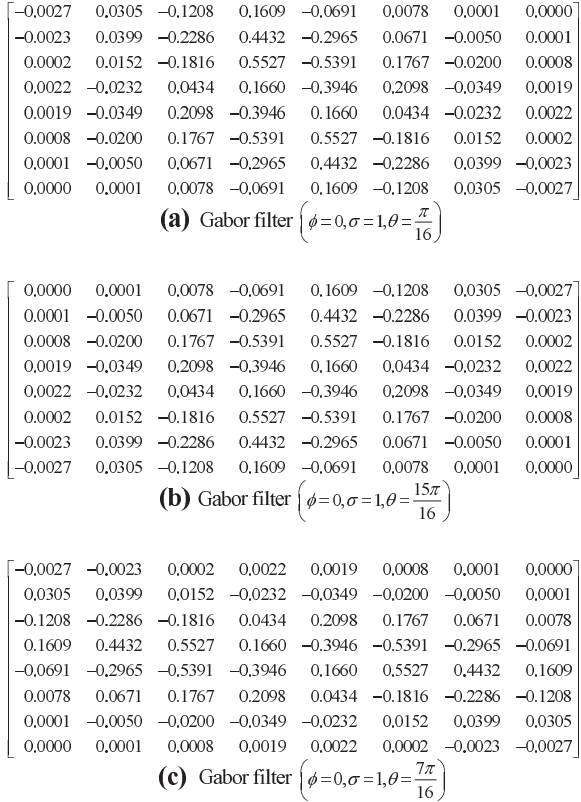}
\caption{Examples of three 2D Gabor filters with different orientations: (a) $\mathbf{G}^{0,1,\pi/16}$, (b) $\mathbf{G}^{0,1,15\pi/16}$, and (c) $\mathbf{G}^{0,1,7\pi/16}$.}
\label{Figure:gabor}
\end{figure}

\textbf{Step 1}: According to the symmetry between Gabor filters $\mathbf{G}^{\phi,\sigma,\theta}$ and $\mathbf{G}^{\phi,\sigma,\pi-\theta}$
(see \textbf{Figure~\ref{Figure:gabor}(a)} and\textbf{~\ref{Figure:gabor}(b)}),
we can merge together the histograms of the subsets of residual images $\mathbf{U}^{\phi,\sigma,\theta}$ and $\mathbf{U}^{\phi,\sigma,\pi-\theta}$.
Specifically, we merge
the histograms $\mathbf{h}^{\phi,\sigma,\theta}_{a,b}$, $\mathbf{h}^{\phi,\sigma,\theta}_{8-a,8-b}$
\big(corresponding to the $(a,b)$th and $(8-a,8-b)$th subsets of $\mathbf{U}^{\phi,\sigma,\theta}$\big)
and the histograms $\mathbf{h}^{\phi,\sigma,\pi-\theta}_{8-a,b}$, $\mathbf{h}^{\phi,\sigma,\pi-\theta}_{a,8-b}$
\big(corresponding to the $(8-a,b)$th and $(a,8-b)$th subsets of $\mathbf{U}^{\phi,\sigma,\pi-\theta}$\big).

\begin{figure}
\centering
\includegraphics[width=1\columnwidth]{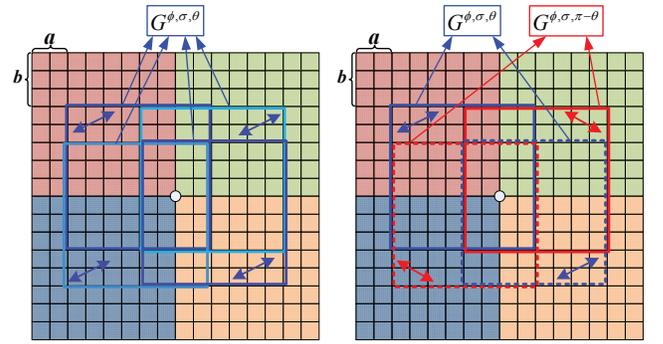}
\caption{Left: The merging method (Step 6 in Section 2) in the original GFR. (The blue windows denote the Gabor filter $\mathbf{G}^{\phi,\sigma,\theta}$. When $\mathbf{G}^{\phi,\sigma,\theta}$ is located at these four positions, four subsets $\mathbf{U}^{\phi,\sigma,\theta}_{a,b}$, $\mathbf{U}^{\phi,\sigma,\theta}_{a,8-b}$, $\mathbf{U}^{\phi,\sigma,\theta}_{8-a,b}$, $\mathbf{U}^{\phi,\sigma,\theta}_{8-a,8-b}$ are computed. The histograms of these four subsets can be merged with the merging method in Step 6 in Section 2.)  Right: The merging method in \textbf{Step} 1 (Section 4) based on the symmetry between $\mathbf{G}^{\phi,\sigma,\theta}$ and $\mathbf{G}^{\phi,\sigma,\pi-\theta}$. (The blue windows denote the Gabor filter $\mathbf{G}^{\phi,\sigma,\theta}$, and the red ones denote $\mathbf{G}^{\phi,\sigma,\pi-\theta}$. When $\mathbf{G}^{\phi,\sigma,\theta}$ and $\mathbf{G}^{\phi,\sigma,\pi-\theta}$ are located at these positions, four subsets $\mathbf{U}^{\phi,\sigma,\theta}_{a,b}$, $\mathbf{U}^{\phi,\sigma,\theta}_{8-a,8-b}$, $\mathbf{U}^{\phi,\sigma,\pi-\theta}_{a,8-b}$, $\mathbf{U}^{\phi,\sigma,\pi-\theta}_{8-a,b}$ are computed. The histograms of these four subsets can be merged with the merging method in \textbf{Step} 1 in Section 4.)}
\label{Figure:mergingMethod1}
\end{figure}

The merging method in \textbf{Step 1} is different from the method used in the DCTR and the original GFR (\textbf{Step 6} in Section~\ref{Sec_2}).
As shown in \textbf{Figure~\ref{Figure:mergingMethod1}},
in the original GFR,  the histograms $\mathbf{h}^{\phi,\sigma,\theta}_{a,b}$, $\mathbf{h}^{\phi,\sigma,\theta}_{8-a,8-b}$,
$\mathbf{h}^{\phi,\sigma,\theta}_{8-a,b}$ and $\mathbf{h}^{\phi,\sigma,\theta}_{a,8-b}$ are from one Gabor residual.
However, in \textbf{Step 1}, we merge the histograms $\mathbf{h}^{\phi,\sigma,\theta}_{a,b}$, $\mathbf{h}^{\phi,\sigma,\theta}_{8-a,8-b}$ and
$\mathbf{h}^{\phi,\sigma,\pi-\theta}_{8-a,b}$, $\mathbf{h}^{\phi,\sigma,\pi-\theta}_{a,8-b}$ that
are from two Gabor residuals $\mathbf{U}^{\phi,\sigma,\theta}$ and $\mathbf{U}^{\phi,\sigma,\pi-\theta}$.
In \textbf{Figure~\ref{Figure:mergingMethod1}}, there is an interesting finding that when computing the subsets whose histograms will be merged according to our method in \textbf{Step 1},
the $8 \times 8$ window of the Gabor filter $\mathbf{G}^{\phi,\sigma,\theta}$ is symmetric with the window of $\mathbf{G}^{\phi,\sigma,\pi-\theta}$
about the boundaries of the $8 \times 8$ pixel blocks (i.e., the blue windows are symmetric with the red windows about the boundaries).

\textbf{Step 2}: Due to the transposition relation between $\mathbf{G}^{\phi,\sigma,\theta}$ and $\mathbf{G}^{\phi,\sigma,\pi/2-\theta}$ (see \textbf{Figure~\ref{Figure:gabor}(a)} and~\textbf{\ref{Figure:gabor}(c)}),
we merge together
the histograms of the $(a,b)$th subset of residual $\mathbf{U}^{\phi,\sigma,\theta}$ and the $(b,a)$th subset of $\mathbf{U}^{\phi,\sigma,\pi/2-\theta}$.

\begin{figure}
\centering
\includegraphics[width=0.5\columnwidth]{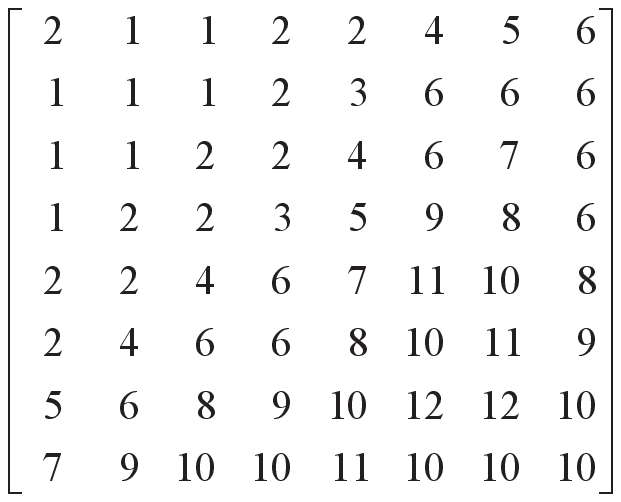}
\caption{The standard JPEG quantization table of quality factor 95.}
\label{Figure:quantable}
\end{figure}

\begin{figure}
\centering
\includegraphics[width=0.5\columnwidth]{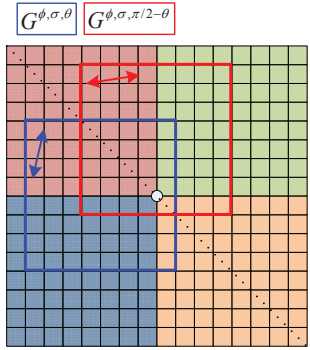}
\caption{The merging method in Step 2 (Section 4) based on the symmetry between $\mathbf{G}^{\phi,\sigma,\theta}$ and $\mathbf{G}^{\phi,\sigma,\pi/2-\theta}$. (The blue window denotes the Gabor filter $\mathbf{G}^{\phi,\sigma,\theta}$, and the red one denotes $\mathbf{G}^{\phi,\sigma,\pi/2-\theta}$. When $\mathbf{G}^{\phi,\sigma,\theta}$ and $\mathbf{G}^{\phi,\sigma,\pi/2-\theta}$ are located at these positions, two subsets $\mathbf{U}^{\phi,\sigma,\theta}_{a,b}$, $\mathbf{U}^{\phi,\sigma,\pi/2-\theta}_{b,a}$ are computed. The histograms of these two subsets can be merged with the merging method in \textbf{Step} 2 in Section 4.)}
\label{Figure:mergingMethod2}
\end{figure}

The merging method in \textbf{Step 2} is based on the argument that a decompressed JPEG image
still somehow preserves the symmetric properties.
Although it is known that the
symmetries of a natures image are broken by the quantization in JPEG compression
due to the rounding operation and the non-symmetric quantization table, we argue that this situation is not serious and it is still reasonable to merge the statistical characteristics according to the spatial diagonal symmetry.
First, for a standard JPEG quantization table (see \textbf{Figure~\ref{Figure:quantable}}), the elements for low-frequency DCT coefficients are symmetric w.r.t. the $8\times8$ block main diagonal, especially for high quality factors.
Second, since most high-frequency DCT coefficients are zeros, they mitigate the impact of non-symmetric elements in the quantization table because actually they produce the same zero value in the dequantization.
From \textbf{Figure~\ref{Figure:mergingMethod2}}, we find that when computing the subsets whose histograms will be merged according to the method in \textbf{Step 2},
the $8 \times 8$ window of the Gabor filter $\mathbf{G}^{\phi,\sigma,\theta}$ is symmetric with the window of $\mathbf{G}^{\phi,\sigma,\pi/2-\theta}$
about the main diagonal (i.e., the blue window is symmetric with the red window about the main diagonal).

In the following, we will demonstrate the reasons for merging histograms in the above two steps
and show the details.

\subsection{Analysis of Merging Method in Step 1}


We find the fact that there exit symmetries between
$\mathbf{G}^{\phi,\sigma,\theta}$ and $\mathbf{G}^{\phi,\sigma,\pi-\theta}$ ($0\leq\theta<\pi$, $\theta \neq \{0,\pi/2\}$):
\begin{equation}
\label{eq_symmetry_pi_Gabors}
\begin{aligned}
&\mathbf{G}^{\phi=0,\sigma,\theta}=&   &\mathrm{fliplr}(\mathbf{G}^{\phi=0,\sigma,\pi-\theta})=& &\mathrm{flipud}(\mathbf{G}^{\phi=0,\sigma,\pi-\theta})&\\
&\mathbf{G}^{\phi=\frac{\pi}{2},\sigma,\theta}=&   &\mathrm{fliplr}(\mathbf{G}^{\phi=\frac{\pi}{2},\sigma,\pi-\theta})=&  -&\mathrm{flipud}(\mathbf{G}^{\phi=\frac{\pi}{2},\sigma,\pi-\theta})&
\end{aligned}.
\end{equation}
Thus, from~(\ref{eq_symmetry_DCT}) and~(\ref{eq_symmetry_pi_Gabors}), we can find the symmetry between $\left|\mathbf{R}^{(\phi,\sigma,\theta)(k,l)}\right|$ and
$\left|\mathbf{R}^{(\phi,\sigma,\pi- \theta)(k,l)}\right|$:
\begin{equation}
\label{eq_symmetry_R_pi&pi-theta}
\begin{aligned}
\left|{R}^{(\phi,\sigma,\theta)(k,l)}_{a,b}\right|  = & \left|{R}^{(\phi,\sigma,\pi-\theta)(k,l)}_{-a,b}\right|\\
\left|{R}^{(\phi,\sigma,\theta)(k,l)}_{a,b}\right|  = & \left|{R}^{(\phi,\sigma,\pi-\theta)(k,l)}_{a,-b}\right|
\end{aligned}.
\end{equation}
According to the definition of projection vector $\mathbf{P}^{(\phi,\sigma,\theta)(k,l)}_{a,b}$ (\ref{eq_definition_projection_GFR}),
we can see the following symmetry by (\ref{eq_symmetry_R_pi&pi-theta}),
\begin{equation}
\label{eq_symmetry_P_pi&pi-theta}
\begin{aligned}
\left|\mathbf{P}^{(\phi,\sigma,\theta)(k,l)}_{a,b}\right|  = & \left|\mathbf{P}^{(\phi,\sigma,\pi-\theta)(k,l)}_{-a,b}\right|\\
\left|\mathbf{P}^{(\phi,\sigma,\theta)(k,l)}_{a,b}\right|  = & \left|\mathbf{P}^{(\phi,\sigma,\pi-\theta)(k,l)}_{a,-b}\right|
\end{aligned}.
\end{equation}
From (\ref{eq_symmetry_P_pi&pi-theta}) and (\ref{eq_symmetry_|P2|_2}), we have
\begin{equation}
\label{eq_symmetry_P_2_pi&pi-theta}
\begin{aligned}
\left|\mathbf{P}^{(\phi,\sigma,\theta)(k,l)}_{a,b}\right|  =  &\left|\mathbf{P}^{(\phi,\sigma,\theta)(k,l)}_{a-8,b}\right|
=  \left|\mathbf{P}^{(\phi,\sigma,\pi-\theta)(k,l)}_{8-a,b}\right|\\
\left|\mathbf{P}^{(\phi,\sigma,\theta)(k,l)}_{a,b}\right|  =  &\left|\mathbf{P}^{(\phi,\sigma,\theta)(k,l)}_{a,b-8}\right|
=  \left|\mathbf{P}^{(\phi,\sigma,\pi-\theta)(k,l)}_{a,8-b}\right|
\end{aligned}.
\end{equation}
Combining the symmetry (\ref{eq_symmetry_P_2_pi&pi-theta}) with the symmetry
$\left|\mathbf{P}^{(\phi,\sigma,\theta)(k,l)}_{a,b}\right|  =  \left|\mathbf{P}^{(\phi,\sigma,\theta)(k,l)}_{8-a,8-b}\right|$ (\ref{eq_symmetry_|P2_2|_3})
, we have
\begin{equation}
\begin{aligned}
\label{eq_symmetry_porposed_1}
\left|\mathbf{P}^{(\phi,\sigma,\theta)(k,l)}_{a,b}\right|  =  &\left|\mathbf{P}^{(\phi,\sigma,\theta)(k,l)}_{8-a,8-b}\right|\\
= &\left|\mathbf{P}^{(\phi,\sigma,\pi-\theta)(k,l)}_{a,8-b}\right| \\
= &\left|\mathbf{P}^{(\phi,\sigma,\pi-\theta)(k,l)}_{8-a,b}\right|
\end{aligned}.
\end{equation}
According to the above symmetry (\ref{eq_symmetry_porposed_1}), the subsets of residual $\mathbf{U}^{\phi,\sigma,\theta}$
obtained with $\mathbf{G}^{\phi,\sigma,\theta}$ and the subsets of residual $\mathbf{U}^{\phi,\sigma,\pi-\theta}$ obtained with $\mathbf{G}^{\phi,\sigma,\pi-\theta}$
can be considered together. As shown in \textbf{Figure~\ref{Figure:mergingMethod1}}, thus, we can merge the histograms of the subsets
corresponding to the positions $(a,b)$, $(8-a,8-b)$
in $\mathbf{U}^{\phi,\sigma,\theta}$ and the subsets corresponding to $(8-a,b)$, $(a,8-b)$ in  $\mathbf{U}^{\phi,\sigma,\pi-\theta}$.
That is,
\begin{equation}
\label{method_merging_1}
\mathbf{h}^{\phi,\sigma,\theta}_{a,b} \leftarrow \mathbf{h}^{\phi,\sigma,\theta}_{a,b} + \mathbf{h}^{\phi,\sigma,\theta}_{8-a,8-b}
+ \mathbf{h}^{\phi,\sigma,\pi-\theta}_{8-a,b} + \mathbf{h}^{\phi,\sigma,\pi-\theta}_{a,8-b}, 0 < \theta < \pi/2
\end{equation}
Note that these indices, $(a,b)$, $(8-a,8-b)$, $(8-a,b)$ and $(a,8-b)$, should stay within $\{0,1,\ldots,7\}\times\{0,1,\ldots,7\}$.
When 
 $(8-a)$ or $(8-b)$ is 8 $\notin \{0,1,\ldots,7\}$,
we can take $\mathrm{mod} 8$ of these indices $\left(\mathrm{mod}(8,8)=0\right)$.

\begin{figure}
\centering
\includegraphics[width=1\columnwidth]{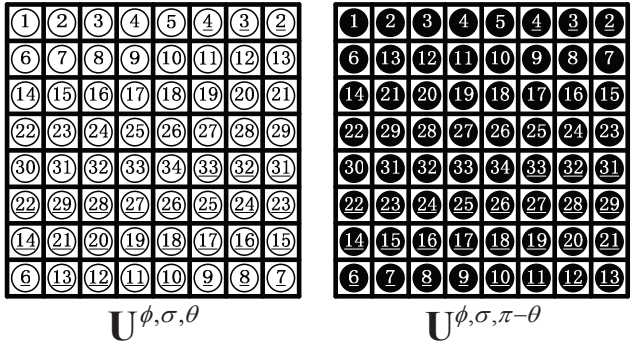}
\caption{The subsets of $\mathbf{U}^{\phi,\sigma,\theta}$ and $\mathbf{U}^{\phi,\sigma,\pi-\theta}$ (a circle denotes a subset ($\mathbf{U}^{\phi,\sigma,\theta}_{a,b}$ or $\mathbf{U}^{\phi,\sigma,\pi-\theta}_{a,b}$), where $(a,b)$ is the circle's location in the $8\times8$ grid). }
\label{Figure:1symmetry}
\end{figure}

For the condition of $\theta \neq \{0, \pi/2\}$,
there are 30 orientations, $L$ scales and 2 phase shifts,
so the number of the Gabor filters is $2\cdot L\cdot 30$.
Without the merging method,
the total dimension of the histograms is
$2\cdot L\cdot 30 \cdot64 \cdot (T+1)$, where $T$ is the histogram threshold.
From \textbf{Figure~\ref{Figure:1symmetry}}, it can be seen that
according to the symmetry between $\mathbf{G}^{\phi,\sigma,\theta}$ and $\mathbf{G}^{\phi,\sigma,\pi-\theta}$,
the dimensions can be reduced to $2 \cdot L \cdot 15 \cdot 34 \cdot (T+1)$ by merging together the histograms of the subsets labeled with the same number (regardless of the color and the underline).

\subsection{Analysis of Merging Method in Step 2}


For $\mathbf{G}^{\phi,\sigma,\theta}$($0\leq\theta\leq\pi/2$), we find that
\begin{equation}
\mathbf{G}^{\phi,\sigma,\theta}  = {\left(\mathbf{G}^{\phi,\sigma,\pi/2-\theta}\right)}^{\emph{T}}, \quad 
\end{equation}
where $(\cdot)^\emph{T}$ indicates the transpose operation.
Thus, according to the symmetry between  $\mathbf{G}^{\phi,\sigma,\theta}$ and $\mathbf{G}^{\phi,\sigma,\pi/2-\theta}$,
the residuals $\mathbf{U}^{\phi,\sigma,\theta}$ and $\mathbf{U}^{\phi,\sigma,\pi/2-\theta}$, which are obtained using the filter $\mathbf{G}^{\phi,\sigma,\theta}$ and its transposed version $\mathbf{G}^{\phi,\sigma,\pi/2-\theta}$, can be
considered together. We can merge together
the histograms of the residuals  $\mathbf{U}^{\phi,\sigma,\theta}$ and $\mathbf{U}^{\phi,\sigma,\pi/2-\theta}$ to further decrease the feature dimension and endow them more robustness.
This idea has been adopted in the PSRM which is one of the most effective steganalysis features in the spatial domain.
As shown in \textbf{Figure~\ref{Figure:mergingMethod2}}, we can merge the histogram $\mathbf{h}^{\phi,\sigma,\theta}_{a,b}$ and $\mathbf{h}^{\phi,\sigma,\pi/2-\theta}_{b,a}$,

\begin{equation}
\mathbf{h}^{\phi,\sigma,\theta}_{a,b} \leftarrow \mathbf{h}^{\phi,\sigma,\theta}_{a,b} + \mathbf{h}^{\phi,\sigma,\pi/2-\theta}_{b,a}, \ 0\leq\theta\leq\pi/4
\end{equation}
where $\mathbf{h}^{\phi,\sigma,\theta}_{a,b}$ is the histogram of the $(a,b)$th subset of residual $\mathbf{U}^{\phi,\sigma,\theta}_{a,b}$,
and $\mathbf{h}^{\phi,\sigma,\pi/2-\theta}_{b,a}$ is the histogram of the $(b,a)$th subset of $\mathbf{U}^{\phi,\sigma,\pi/2-\theta}_{b,a}$.
Note that
the indices of these two subsets, $\mathbf{U}^{\phi,\sigma,\theta}_{a,b}$ and $\mathbf{U}^{\phi,\sigma,\pi/2-\theta}_{b,a}$,
are transposed to avoid mixing up different statistical characteristics.
This is because when the filter is transposed,
the phase-aware statistics of the filtered image
are transposed accordingly.

According to the symmetry between $\mathbf{G}^{\phi,\sigma,\theta}$ and $\mathbf{G}^{\phi,\sigma,\pi/2-\theta}$, the dimensions can be decreased furthermore. For the condition of $\theta \neq \{0, \pi/2\}$,
the feature vector of $2 \cdot L \cdot 15 \cdot 34 \cdot (T+1)$ dimensions can be reduced to $2 \cdot L \cdot 8 \cdot 34 \cdot (T+1)$.
For the condition of $\theta = \{0, \pi/2\}$,
the $2 \cdot L \cdot 2 \cdot 25 \cdot (T+1)$ dimensions can be reduced to $2 \cdot L \cdot 1 \cdot 25 \cdot (T+1)$.

To sum up, with our proposed merging method in Section~\ref{Sec_4}, the dimension of the improved GFR features (GFR-GSM) is $594 \cdot L \cdot (T+1)$\footnote{$2 \cdot L \cdot 8 \cdot 34 \cdot (T+1) +  2 \cdot L \cdot 1 \cdot 25 \cdot (T+1)$ = $594 \cdot L \cdot (T+1)$}.
If the number of scales $L = 4$ and the histogram threshold $T = 4$ are the same as in the original GFR~\cite{GFR}, the dimensions are reduced to 11880.
From the experiments in Section~\ref{Sec_6}, when compared with the 17000-dimensional GFR, the 11880-dimensional GFR-GSM$_{4}$ (the subscript 4 denotes the number of scales $L = 4$) can achieve better detection performance with smaller dimensions.

\section{Proposed Weighted Histogram Method}
\label{Sec_5}
No matter in the GFR or in the DCTR, all the absolute values of residuals are quantized to the integer values before computing the phase-aware histograms.
Specifically, in the GFR, the residual $\left|\mathbf{U}^{\phi,\sigma,\theta}\right| = |{u}^{\phi,\sigma,\theta}_{kl}|$ is divided by the quantization step $q$ and quantized with a quantizer $Q_T$ with $T$ + 1 centroids $Q=\{0, 1, \ldots T\}$,

\begin{equation}
Q_{T}(|{u}^{\phi,\sigma,\theta}_{kl}|/q) =
\mathrm{trunc}_{T} \left( \mathrm{round} \left( |{u}^{\phi,\sigma,\theta}_{kl}|/q \right) \right),
\end{equation}
where $\mathrm{round}(\cdot)$ denotes the rounding operation, and $\mathrm{trunc}_{T}(\cdot)$ denotes the truncation with the threshold $T$.
The values of residuals are mapped to the integers ($Q$) through the above quantization.
Although the quantization can curb the dimensionality of the feature space, it inevitably leads to loss of useful information.
With the quantization,
the residual samples, which are quantized to the same centroid, are always
located in different positions within the same interval.
This means the slight changes in residual samples caused by embedding may be left out,
which may affect the detection accuracy.

In this section, we associate a residual sample with a Gaussian function
and use the integrals over all quantization intervals as the weights that will be accumulated into the corresponding histogram bins.
This method refers to the soft voting scheme that has been used in other fields of machine learning~\cite{soft}.
This histogram method can also be applied to other histogram features, such as the PSRM, the PHARM and the DCTR.

Each residual sample is associated with a Gaussian function centered at ${u}_{kl}$, $\mathbf{Gauss}({u}_{kl},\sigma_{H}^2)$,
where ${u}_{kl}$ is the value of the residual sample and $\sigma_{H}$ is an important parameter that needs to be adjusted carefully.
In our method, there are $2T+1$ centroids $\{-Tq, \ldots, -q, 0, q, \ldots, Tq\}$.
The interval $I_i$ w.r.t. the centroid $i$ can be expressed as:
\begin{equation}
I_{i} = \left\{
\begin{array}{lll}
\left(-\infty, \right.    &\left. (-T+1/2)q \right]   &i = -T,\\
\left((i-1/2)q, \right.   &\left. (i+1/2)q  \right]   &i = \{-T+1, \ldots, -1\},\\
\left(-1/2\,q,     \right.  &\left. 1/2\,q      \right)   &i = 0,\\
\left[(i-1/2)q,   \right. &\left. (i+1/2)q  \right)   &i = \{1, \ldots, T-1\},\\
\left[( T - 1/2 )q,\right.&\left. \infty    \right)   &i = T.
\end{array} \right.
\end{equation}

As shown in \textbf{Figure~\ref{Figure:weighted_histogram}}, $P_{i}$ is the integral of $\mathbf{Gauss}$ over the interval $I_i$,
and it can be computed as:

\begin{figure}
\centering
\includegraphics[width=\columnwidth]{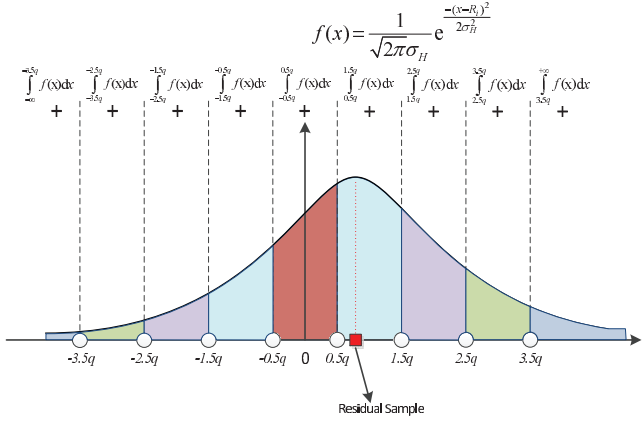}
\caption{Our weighted voting scheme for histogram computation.}
\label{Figure:weighted_histogram}
\end{figure}

\begin{equation}
P_{i} = \left\{
\begin{array}{l}
\int_{-\infty}^{(-T+1/2)q} \frac{1}{\sqrt{2\pi}\sigma_{H}}\mathrm{{exp}}{\left(-({x}-{u}_{kl})^2/\sigma_{H}^2\right)} \, d   x \\ i = -T,\\ \\
\int_{(i-1/2)q}^{(i+1/2)q} \frac{1}{\sqrt{2\pi}\sigma_{H}}\mathrm{{exp}}{\left(-({x}-{u}_{kl})^2/\sigma_{H}^2\right)} \, d
x  \\
i = \{-T+1, \ldots, T-1\},\\ \\
\int_{(T-1/2)q}^{\infty} \frac{1}{\sqrt{2\pi}\sigma_{H}}\mathrm{{exp}}{\left(-({x}-{u}_{kl})^2/\sigma_{H}^2\right)} \, d
x \\ i = T.\\
\end{array} \right..
\end{equation}

In the original GFR, if $|{u}_{kl}|$ falls into the quantization interval $I_i$, we add a 1 to the histogram bin $b_{i}$.
In our method, however,
the weights $P_{i}$ are accumulated into the corresponding histogram bins $b_{i}$.
For $T = 2$, we add $P_{-2}$ to the histogram bin $b_{-2}$ corresponding to the interval $I_{-2}=(-\infty, -1.5q)$,
while adding $P_{-1}$, $P_{0}$, $P_{1}$, $P_{2}$ to the histogram bins $b_{-1}$, $b_{0}$, $b_{1}$, $b_{2}$, respectively.
After computing the weights of all intervals, $P_{i}$ is merged with $P_{-i}$ due to the sign-symmetry
\begin{equation}
P_{i} = \left\{
\begin{array}{lcl}
P_{i} + P_{-i} & & i = \{1,2, \ldots, T\},\\
P_{0} & & i = 0.
\end{array}\right.
\end{equation}
Consequently, the final weighted histogram consists of $T+1$ bins ($b_{i}$, $i=0,1, \ldots, T$).
The complete weighted histogram $\mathbf{h}_{\mathrm{WEIGHT}}$ is computed by summing the contributions of all the samples in the residual image
\begin{equation}
\mathbf{h}_{\mathrm{WEIGHT}}(i) = \sum_{k,l} \int_{I_{i} \bigcup I_{-i}} \frac{1}{\sqrt{2\pi}\sigma_{H}}\mathrm{{exp}}{\left(-({x}-{u}_{kl})^2/\sigma_{H}^2\right)} \ d \,  x,
\end{equation}
where $i \in \{0,1, \ldots ,T\}$.

There are two main differences between our histogram method and the conventional histogram method in the GFR.
First, in our method, the contribution of a residual sample to a bin is a real value rather than a constant value 1 in the conventional method.
Second, in our method, a residual sample contributes to all bins rather than only one bin in the conventional method.

\begin{figure}
\centering
\includegraphics[width=\columnwidth]{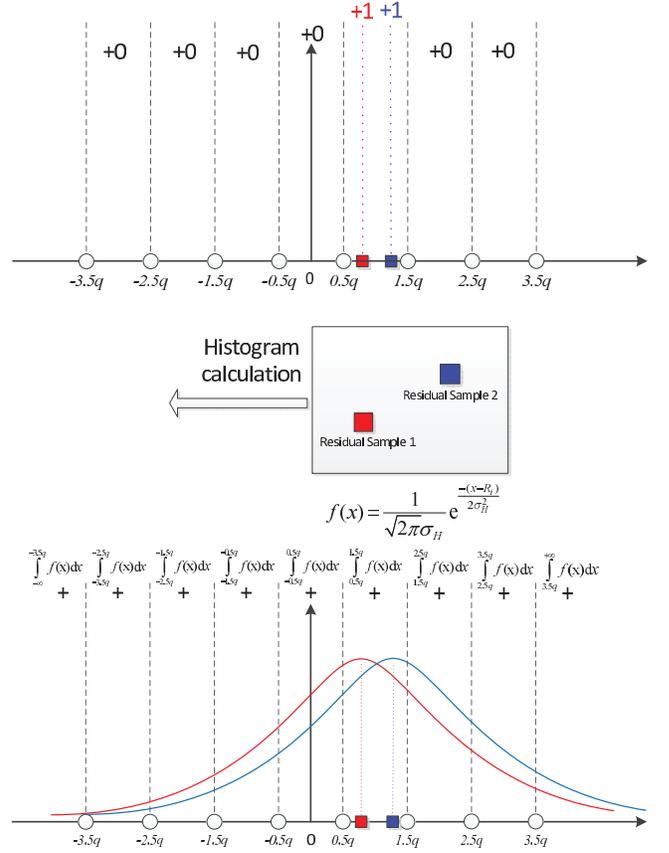}
\caption{The difference between the weighted histogram and the conventional histogram.}
\label{Figure:histogram_compare}
\end{figure}

Our histogram method takes into consideration the positions of residual values in the quantization interval, thus reflecting the slight shift in the interval.
We take \textbf{Figure~\ref{Figure:histogram_compare}} as an example. We can see that
residual sample 1 and residual sample 2 with different values are in the same interval,
even with the same distance to the centroid.
The conventional histogram method in the GFR can not differentiate them.
However, the integral values obtained from the Gaussian function of residual sample 1 are different from residual sample 2. These integral values, as the weights, are accumulated into the histogram, so these two residual samples have different influence on the weighted histogram in our method.

\section{Experiments}
\label{Sec_6}
This section is organized as follows.
In section 6.1, the parameters are discussed for  better detection performance.
In section 6.2, experimental results show the advantages of the proposed steganalysis features.
In the experiments, 10000 $512 \times 512$ grayscale images from BOSSbase are converted into JPEG images with quality factors 75 and 95 as cover images. The advanced adaptive steganographic schemes UED-JC and J-UNIWARD are used to generate stego images with different embedding rates.

The detection accuracy is quantified using the minimal total error probability under equal priors $P_{\mathrm{E}} = \mathrm{min}_{P_{\mathrm{FA}}}\frac{1}{2}(P_{\mathrm{FA}}+P_{\mathrm{MD}})$, where $P_{\mathrm{FA}}$ and
$P_{\mathrm{MD}}$ are the false-alarm and missed-detection probabilities.
The FLD ensemble classifier~\cite{EN} is used in the training and testing stages.
The $\overline{P}_{\mathrm{E}}$ is averaged over ten random 5000/5000 database splits.

\subsection{Parameter Setting}
\subsubsection{Number of Scales of 2D Gabor Filter}
\label{Sec_6_1_1}

%
In this paper, the parameters of 2D Gabor filters $\phi$ and $\theta$ are the same as in the original GFR.
If the scale parameter $\sigma$ of 2D Gabor filters is the same as in the original GFR ($\sigma = 0.5, 0.75, 1, 1.25$),
there are 4 scales and the total dimension of the proposed GFR-GSM$_{4}$ (or GFR-GW$_{4}$) is 11880.
Since our histogram merging method in Section~\ref{Sec_4} reduces the dimensions dramatically,
we can increase the number of scales by adding $\sigma$ = 1.5, 1.75 to improve the accuracy.
This gives our final steganalysis feature set GFR-GW$_{6}$ the dimension of 17820, which is close to the dimension of the original GFR.
The setting of the quantization step $q$ is related to the value of the scale parameter $\sigma$. For the scale parameter in this paper $\sigma$ = 0.5, 0.75, 1, 1.25, 1.5, 1.75,
by referring to the literature~\cite{GFR}, $q$ is set as $q = 2, 4, 6, 8, 10, 12$, respectively when the quality factor is 75, and $q = 0.5, 1, 1.5, 2, 2.5,3$, respectively when the quality factor is 95.

\subsubsection{Parameter $\sigma_{H}$ in Weighted Histogram Method}
\begin{figure}
\centering
\includegraphics[width=0.8\columnwidth]{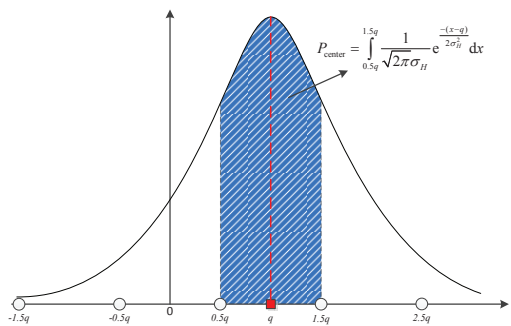}
\caption{
The Gaussian function is centered at the centroid of a quantization interval and the integral over this quantization interval is $P_{\textrm{center}}$.}
\label{Figure:Pcenter}
\end{figure}

To better determine the value of the parameter $\sigma_{H}$, we first introduce a new parameter $P_{\textrm{center}}$.
As shown in \textbf{Figure~\ref{Figure:Pcenter}},
when the Gaussian function is centered at the centroid of a quantization interval $I_i$, the integral over the interval $I_i$ is called
$P_{\textrm{center}}$, $0< P_{\textrm{center}} < 1$.
The value of $P_{\textrm{center}}$ depends on the parameter $\sigma_{H}$ and the quantization step $q$.
In \textbf{Table~\ref{tab:para1}} and \textbf{Table~\ref{tab:para2}}, the effects of the parameter $P_{\textrm{center}}$ on detection accuracy are shown for J-UNIWARD with 0.2 bpnzac payload for quality factors 75 and 95.
From \textbf{Table~\ref{tab:para1}} and \textbf{Table~\ref{tab:para2}}, it can be seen that for the scale parameter $\sigma = 1$ and quality factors 75 and 95,
the best detection accuracy is achieved when $P_{\textrm{center}}$ is equal to 0.75.
For each experiment, since the scale $\sigma$ and the quality factor are fixed, the quantization step $q$ is fixed and $P_{\textrm{center}}$ is only decided by $\sigma_{H}$.
Thus, we maintain that in the case of various scales $\sigma$ and quality factors, $\sigma_{H}$ is always set to make $P_{\textrm{center}}$ equal to $0.75$ for better performance.

\begin{table}
  \caption{The effect of the parameter $P_{\textrm{center}}$ (decided by $\sigma_{H}$) on detection accuracy for $\sigma = 1$ and quality factor 75 ($q = 6$).}
  \centering
  \label{tab:para1}
  \begin{tabular}{cccccc}
    \toprule
    $\sigma_{H}$ &1.8182 &2.0833 &2.3438 &2.6087 &2.8846\\
    \midrule
    $P_{\textrm{center}}$ &0.9 &0.85 &0.8 &0.75 &0.7\\
    \midrule
    $\overline{P}_{\mathrm{E}}$       & 0.3160 & 0.3151 & 0.3149 & \textbf{0.3134} & 0.3142 \\
   \bottomrule
\end{tabular}

%

\end{table}

\begin{table}
  \caption{The effect of the parameter $P_{\textrm{center}}$ (decided by $\sigma_{H}$) on detection accuracy for $\sigma = 1$ and quality factor 95 ($q = 1.5$).}
  \centering
  \label{tab:para2}
  \begin{tabular}{cccccc}
    \toprule
    $\sigma_{H}$ &0.4545 &0.5208 &0.5859 &0.6522 &0.7212\\
    \midrule
    $P_{\textrm{center}}$ &0.9 &0.85 &0.8 &0.75 &0.7\\
    \midrule
    $\overline{P}_{\mathrm{E}}$       & 0.4307 & 0.4307 & 0.4305 & \textbf{0.4297} & 0.4311 \\
   \bottomrule
\end{tabular}
\end{table}

%

\subsection{Experimental Results}

\begin{table*}
  \caption{Difference between GFR and our proposed features.}
  \centering
  \label{tab:1}
  \begin{tabular}{ccccc}
    \toprule
    features        &number of scales of   &dimension &using new histogram merging method  &using our weighted histogram method\\
                    &2D Gabor filter       &          &described in Section 4    &described in Section 5\\
    \midrule
    GFR             &4                     & 17000     & $\times$                 & $\times$              \\
    GFR-GSM$_{4}$     &4                     & 11880     & $\surd$                  & $\times$              \\
    GFR-GW$_{4}$     &4                     & 11880     & $\surd$                  & $\surd$               \\
    GFR-GW$_{6}$     &6                     & 17820     & $\surd$                  & $\surd$               \\
   \bottomrule
\end{tabular}
\end{table*}

Numerous experiments are conducted to demonstrate the effectiveness of the proposed methods.
\textbf{Table~\ref{tab:1}} demonstrates the characteristics of our three proposed feature sets and shows the
difference between the GFR and our feature sets.

From \textbf{Table~\ref{tab:results}}, compared to the 17000-dimensional GFR, the GFR-GSM$_{4}$ with 11880 dimensions, which exploits the proposed histogram merging method, has better detection performance for different steganographic algorithms and embedding rates. This demonstrates that our merging method not only reduces more dimensions but also improves the detection accuracy.
Next, the GFR-GW$_{4}$ using our weighted histogram method achieves better detection accuracy than the GFR-GSM$_{4}$ because the weighted histograms are more sensitive to the small changes than the conventional histograms.
In addition, the detection accuracy of the GFR-GW$_{6}$ is higher than the GFR-GW$_{4}$. This is because the extraction of features from more scales can enhance the diversity and effectiveness of the features. In contrast to 17000-dimensional GFR,
the 17820-dimensional GFR-GW$_{6}$ significantly improves the detection performance
regardless of quality factors, embedding algorithms and embedding rates.
The maximum performance improvement of the GFR-GW$_{6}$ over the original GFR is close to $2.5\%$ for the UED-JC for quality factor 75 with an embedding rate of 0.1 bpnzac.

\begin{table*}
  \caption{Detection error $\overline{P}_{\mathrm{E}}$ for UED-JC and J-UNIWARD for quality factors 75 and 95 when steganalyzed with PHARM, GFR, and our three feature sets.}
  \centering
  \label{tab:results}
  \begin{tabular}{lcccccc}
    \toprule
     J-UNI, QF 75 &0.05 bpnzac &0.1 bpnzac &0.2 bpnzac &0.3 bpnzac &0.4 bpnzac\\
    \midrule
    12600D PHARM       & 0.4741$\pm$0.0023 & 0.4294$\pm$0.0034 & 0.3164$\pm$0.0042 & 0.2099$\pm$0.0036& 0.1271$\pm$0.0024 \\
    17000D GFR         & 0.4638$\pm$0.0028 & 0.4089$\pm$0.0016 & 0.2866$\pm$0.0025 & 0.1786$\pm$0.0033 & 0.1028$\pm$0.0028 \\
    11880D GFR-GSM$_{4}$  & 0.4623$\pm$0.0031 & 0.4058$\pm$0.0027 & 0.2824$\pm$0.0032 & 0.1743$\pm$0.0025 & 0.0990$\pm$0.0023 \\
    11880D GFR-GW$_{4}$   & 0.4586$\pm$0.0023 & 0.3994$\pm$0.0028 & 0.2722$\pm$0.0040 & 0.1651$\pm$0.0024 & 0.0908$\pm$0.0029 \\
    17820D GFR-GW$_{6}$   & 0.4575$\pm$0.0024 & 0.3975$\pm$0.0026 & 0.2685$\pm$0.0040 & 0.1628$\pm$0.0038 & 0.0895$\pm$0.0023 \\
  \bottomrule
\end{tabular}

\begin{tabular}{lccccc}
\\
\\
    \toprule
     UED-JC, QF 75 &0.05 bpnzac &0.1 bpnzac &0.2 bpnzac &0.3 bpnzac &0.4 bpnzac\\
    \midrule
    12600D PHARM        & 0.4217$\pm$0.0017 & 0.3295$\pm$0.0034 & 0.1694$\pm$0.0030 & 0.0798$\pm$0.0029 & 0.0346$\pm$0.0022 \\
    17000D GFR & 0.4090$\pm$0.0041 & 0.3124$\pm$0.0038 & 0.1547$\pm$0.0035 & 0.0707$\pm$0.0022 & 0.0304$\pm$0.0019 \\
    11880D GFR-GSM$_{4}$   & 0.4070$\pm$0.0040 & 0.3071$\pm$0.0032 & 0.1487$\pm$0.0023 & 0.0660$\pm$0.0021 & 0.0271$\pm$0.0015 \\
    11880D GFR-GW$_{4}$   & 0.3962$\pm$0.0022 & 0.2943$\pm$0.0030 & 0.1369$\pm$0.0037 & 0.0611$\pm$0.0025 & 0.0248$\pm$0.0014 \\
    17820D GFR-GW$_{6}$  & 0.3920$\pm$0.0035 & 0.2870$\pm$0.0032 & 0.1336$\pm$0.0037 & 0.0585$\pm$0.0025 & 0.0231$\pm$0.0012 \\
  \bottomrule
\end{tabular}

\begin{tabular}{lccccc}
\\
\\
    \toprule
     J-UNI, QF 95 &0.05 bpnzac &0.1 bpnzac &0.2 bpnzac &0.3 bpnzac &0.4 bpnzac\\
    \midrule
    12600D PHARM      & 0.4945$\pm$0.0022 & 0.4821$\pm$0.0023 & 0.4378$\pm$0.0035 & 0.3803$\pm$0.0038 & 0.3090$\pm$0.0033 \\
    17000D GFR         & 0.4932$\pm$0.0023 & 0.4751$\pm$0.0020 & 0.4232$\pm$0.0042 & 0.3506$\pm$0.0038 & 0.2703$\pm$0.0056 \\
    11880D GFR-GSM$_{4}$   & 0.4910$\pm$0.0025 & 0.4738$\pm$0.0020 & 0.4202$\pm$0.0034 & 0.3477$\pm$0.0045 & 0.2661$\pm$0.0032 \\
    11880D GFR-GW$_{4}$    & 0.4899$\pm$0.0019 & 0.4715$\pm$0.0034 & 0.4157$\pm$0.0025 & 0.3421$\pm$0.0037 & 0.2611$\pm$0.0042 \\
    17820D GFR-GW$_{6}$    & 0.4897$\pm$0.0020 & 0.4709$\pm$0.0017 & 0.4153$\pm$0.0026 & 0.3417$\pm$0.0025 & 0.2583$\pm$0.0034 \\
  \bottomrule
\end{tabular}

\begin{tabular}{lccccc}
\\
\\
    \toprule
     UED-JC, QF 95 &0.05 bpnzac &0.1 bpnzac &0.2 bpnzac &0.3 bpnzac &0.4 bpnzac\\
    \midrule
    12600D PHARM      & 0.4799$\pm$0.0018 & 0.4482$\pm$0.0035 & 0.3698$\pm$0.0038 & 0.2789$\pm$0.0034 & 0.1966$\pm$0.0020 \\
    17000D GFR        & 0.4695$\pm$0.0028 & 0.4325$\pm$0.0028 & 0.3420$\pm$0.0037 & 0.2486$\pm$0.0030 & 0.1647$\pm$0.0031 \\
    11880D GFR-GSM$_{4}$   & 0.4682$\pm$0.0018 & 0.4297$\pm$0.0029 & 0.3380$\pm$0.0025 & 0.2413$\pm$0.0040 & 0.1602$\pm$0.0024 \\
    11880D GFR-GW$_{4}$    & 0.4663$\pm$0.0021 & 0.4258$\pm$0.0036 & 0.3299$\pm$0.0050 & 0.2345$\pm$0.0038 & 0.1551$\pm$0.0036 \\
    17820D GFR-GW$_{6}$    & 0.4654$\pm$0.0020 & 0.4243$\pm$0.0031 & 0.3257$\pm$0.0039 & 0.2334$\pm$0.0029 & 0.1521$\pm$0.0033 \\
  \bottomrule
\end{tabular}
\end{table*}


\section{Improving features via CNN}
\label{Sec_7}
Recently, the convolutional neural networks (CNNs) have attracted much attention in the field of image steganalysis
due to their great achievements in the computer vision.
And several promising CNN architectures have been proposed
to show the great potential of the CNN-based steganalysis~\cite{Tan, Qian, Xu_SPL, Xu_IH16, Sedighi, Chenmo, Ni}.
From these network architectures,
we find that the modules of CNNs for steganalysis are much or less similar to the processes
for the conventional feature-based steganalysis.
Like the feature-based detector,
the network equipped with the high-pass filtering (HPF) layer first transforms the input images to the residuals so as to strengthen the stego signal.
The absolute activation (ABS) layer is proposed to leverage the sign symmetry
which is commonly used in traditional steganalytic schemes.
The phase-spilt layer forces the Chen's PNet and VNet~\cite{Chenmo} to take into account the knowledge of JPEG phase
which is originally employed in the JPEG-phase-aware features.
The histogram layer is implemented in Sedighi's network~\cite{Sedighi} to simulate the formation of histograms in PSRM.
These observations suggest that the design of a CNN detector benefits from the insights and experiences gained from conventional feature-based steganalysis.
To further make use of the domain knowledge,
a novel CNN architecture, with proper initialization, is elaborated to duplicate the steganalytic scheme
with GFR-GW features and FLD-ensemble.

The primary advantages of this architecture can be concluded as follows.
First, the proposed network is capable of optimizing the design of filters in phase-aware features.
Within our CNN framework, we convolve the kernels in the HPF layer with the ones in the convolutional layer to form the kernels which can be used to generate residuals.
Since the kernel weights in the convolutional layer are learned during training,
we have an opportunity to obtain the optimized kernels which can be adopted to improve the performance of the conventional JPEG steganalysis.
Second, with the knowledge of GFR-GW features and FLD-ensemble, our network initially works well, thus facilitating the convergence of the network.
And the batch normalization (BN) layer is not needed in our network since the CNN training with a good initialization is not easy to fall into poor local minima.
Third, our network is not deep, so it is possible to further modify the CNN architecture  by
increasing more convolutional layers.

\begin{figure}
\centering
\includegraphics[width=1\columnwidth]{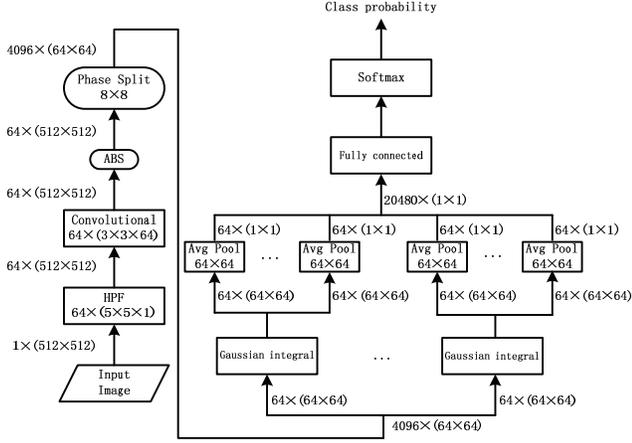}
\caption{
The proposed CNN architecture.}
\label{Figure:CNN}
\end{figure}

The key to our CNN framework is how to model the feature-based detector.
To better understand our architecture, we first briefly review the
computational procedures of the detector with GFR-GW and FLD-ensemble, including
(Step 1) filtering using 2D Gabor filters; (Step 2) splitting by the JPEG phases; (Step 3) computing the weighted histograms using Gaussian-integral; (Step 4) merging based on symmetries; (Step 5) classification with FLD-ensemble.
Next, we will describe in detail the modules in our network which can simulate these procedures well (See \textbf{Figure~\ref{Figure:CNN}}).

($\mathcal{A}$)
In our framework, the HPF layer and the convolutional layer are combined to represent the process of Gabor filtering (Step 1).
In the HPF layer we employ 64 $5\times5$ Gabor filters $g$ (16 orientations, 2 scales and 2 phases) as the high-pass filters, whose parameters are fixed during the training.
In the convolutional layer we use 64 $3\times3\times64$ kernels.
Instead of the random initialization, the $K$th convolutional kernel $f_K \in \mathbb{R}^{3\times3\times64}$
is initialized as
\begin{equation}
  f_K(:,:,k) = \left\{
  \begin{aligned}
   &\left[
\begin{matrix}
0&0&0\\
0&1&0\\
0&0&0
\end{matrix}
\right],&  &k = K& \\
   &\left[
\begin{matrix}
0&0&0\\
0&0&0\\
0&0&0
\end{matrix}
\right],&  &k \neq K&
  \end{aligned}
  \right..
\end{equation}
Due to the fact that convolution is associative, convolution of $g$ with the kernel $f_K$ is equivalent to a $7\times7$ kernel whose central $5\times5$ portion is the $K$th $5\times5$ Gabor filter surrounded by zeros.
Thus, with the initialized parameters, the output feature maps of the convolutional layer is the same as the residual images generated by convolving with 64 $5\times5$ Gabor filters.
Since the parameters of the convolution layer are updated during training, the optimized kernels can help to obtain more suitable filters to form residuals in JPEG-phase-aware steganalysis features.

($\mathcal{B}$)
The phase-split layer is inserted to split the output of the ABS layer into 64 groups according to their JPEG phases (Step 2).
The phase-split layer in our network is the same as  the one in Chen's PNet and VNet.
The difference is that the features generated from all phase groups will be merged together in the fully-connected layer (Step 4).
The weights from symmetric phase groups are initialized with the same value to taken into account the symmetrization utilized in the GFR-GW.
Note that, since the size of Gabor filter in HPF layer is $5\times5$, the merging scheme in our network is different from the GFR-GW where the $8\times8$ filters are used.

($\mathcal{C}$)
The Gaussian-integral layer, followed by global averaging pooling layer, is placed to implement the weighted histograms of subimages in the GFR-GW (Step 3).
In~\cite{Sedighi}, Sedighi's histogram layer is used to simulate the conventional histogram using the mean-shifted Gaussian kernels.
But our Gaussian-integral layer is employed to compute the weighted histogram. The weights are computed as the integrals of a Gaussian function over
different intervals, which can be represented by using Gaussian activations.
To match the 5-bin weighted histogram in GFR-GW, 5 Gaussian-integral layers are used to compute the
histogram bins $B(i)$.
For an $M \times N$ feature map $U = u_{kl}$, the value of $B(i)$, taking into account the sign-symmetry, can be computed as:

\begin{equation}
B(i) = \sum_{k=1}^{M}\sum_{l=1}^{N}
\int_{I_{i} \cup -I_{i}} \frac{1}{\sqrt{2\pi}\sigma_{H}}\mathrm{{exp}}{\left(-({x}-{u}_{kl})^2/\sigma_{H}^2\right)} \, dx
\end{equation}
where
\begin{equation}
  I_{i} = \left\{
  \begin{aligned}
  &[0, 0.5q),& &i=0& \\
  &[0.5q, 1.5q),& &i=1& \\
  &[1.5q, 2.5q),& &i=2& \\
  &[2.5q, 3.5q),& &i=3& \\
  &[3.5q, +\infty),& &i=4& \\
  \end{aligned}
  \right..
\end{equation}
All computed histograms will be concatenated and passed to the fully-connected layers for classification.
During back propagation, the gradient of the loss function $L$ with respect to each element of the feature maps $u_{kl}$ will be computed as:

\begin{equation}
\begin{aligned}
\frac{\partial L}{\partial u_{kl}} = &\sum_{i=0}^{4}\frac{\partial L}{\partial B(i)}\frac{\partial B(i)}{\partial u_{kl}}\\
= &\sum_{i=0}^{4}\frac{\partial L}{\partial B(i)}\frac{\partial \int_{I_{i} \cup -I_{i}} \frac{1}{\sqrt{2\pi}\sigma_{H}}\mathrm{{exp}}{\left(-({x}-{u}_{kl})^2/\sigma_{H}^2\right)} \, dx}{\partial u_{kl}}\\
= &\sum_{i=0}^{4}
\frac{\partial L}{\partial B(i)}
\int_{I_{i} \cup -I_{i}}
\frac{\partial \frac{1}{\sqrt{2\pi}\sigma_{H}}\mathrm{exp}\left(-({x}-{u}_{kl})^2/\sigma_{H}^2\right)}{\partial u_{kl}}
\, dx\\
= &\sum_{i=0}^{4}\frac{\partial L}{\partial B(i)}\frac{{f(b_{i})} - {f(a_{i})} + {f(-a_{i})} - {f(-b_{i})}}{-\sqrt{2\pi}\sigma_{H}}
\end{aligned}
\end{equation}
where $f(x) = \mathrm{{exp}}{\left(-({x}-{u}_{kl})^2/\sigma_{H}^2\right)}$, $a_{i}$ and $b_{i}$ are the lower and upper boundaries of $I_{i}$, respectively.
The difference between Sedighi's net and ours is that the output of Sedighi's histogram layer is the value of a Gaussian function while ours is the Gaussian integral.

($\mathcal{D}$)
The fully-connected layer and the softmax layer are implemented to model the FLD-ensemble. In the fully-connected layer the number of node is the same as the number of chosen FLDs,
and the weights are initialized with the already-trained FLD-ensemble.
For those unselected features, the weights are set to zero.

With above well-designed modules, the network can duplicate the scheme with GFR-GW and FLD-ensemble. The trained convolutional kernels are convolved with the fixed kernels in HPF layer to generate 64 $7\times7$ kernels which maybe more proper filters than Gabor filters used to  generate residuals.

\section{Conclusion}
\label{Sec_8}
In this paper, we modify the original GFR features for better detection performance.
There are two main contributions in this paper.
First, according to the symmetries between different Gabor filters, we merge the histograms in a special way,
thus compactifying the features furthermore while improving the detection accuracy.
Second, our weighted histogram method is more sensitive to the small changes in residuals, simply placing a Gaussian on each of the residual samples and
using the integrals over quantizing intervals.
With these two improvements, the proposed GFR-GW$_6$ with similar dimensions is more powerful than the original GFR.
We also propose a CNN to duplicate the feature-based detector with GFR-GW and FLD-ensemble in order to train better filters for residuals in JPEG-phase-aware features.

The future work will focus on the following several aspects. First, we can merge the DCTR features according to the transposition relation between different DCT kernels to reduce the dimensions furthermore.
Second, in our weighted histogram method, the integral values of the Gaussian function are computed via the MATLAB command 'normcdf', which is expensive in computation time. So we can first save the table of integrals in the memory and then use the method of table look-up to make our histogram method more practically efficient.
Third, when computing the histograms using a weighted voting scheme, the weight can be calculated with other strategies.
Fourth, some parameters in our methods, such as $\sigma_{H}$, are tuned thanks to preliminary experiments done on BOSSbase, which may lead to a kind of overfitting on the BOSSbase. So we will further validate the effectiveness of the parameters on other image bases.
Fifth, as a universal feature set, the GFR-GW$_6$ can also be modified to be a selection-channel-aware version with the method in~\cite{SCA-JPEG} to detect adaptive steganography more accurately.
Sixth, like the GPU-version of steganalysis features (e.g., GPU-PSRM~\cite{GPU-PSRM}, GPU-SRM and GPU-DCTR~\cite{GPU-SRM}), our proposed features can also be implemented on the GPU device to make them more efficient.
Although the Gabor filters is not separable, it can be decomposed using the SVD method to accelerate the filtering~\cite{GPU-Gabor}. So it is not very difficult to implement our features on a GPU.

\begin{acks}
This work was supported by the NSFC under U1536105 and U1636102, and
National Key Technology R\&D Program under 2014BAH41B01, 2016YFB0801003 and 2016QY15Z2500.
\end{acks}
\bibliographystyle{ACM-Reference-Format}
\balance
\bibliography{xiaIH17}
\end{document}